\documentclass{article}
\usepackage[utf8]{inputenc}
\usepackage{amsfonts}
\usepackage{amsmath}
\usepackage{amssymb}
\usepackage{fullpage}
\usepackage{hyperref}
\usepackage{setspace}
\usepackage{graphicx}
\usepackage{geometry}
\usepackage{amsthm}
\usepackage{multirow}
\usepackage{float}
\usepackage[affil-it]{authblk}
\newtheorem{theorem}{Theorem}[section]

\usepackage{xcolor}

\newtheorem{definition}{Definition}
\begin{document}
\title{Model Identifiability for Bivariate Failure Time Data with Competing Risks: Parametric Cause-specific Hazards and Non-parametric Frailty}
\author[1]{Biswadeep Ghosh}
\author[1]{Anup Dewanji}
\author[2]{Sudipta Das}
\affil[1]{Applied Statistics Unit, Indian Statistical Institute}
\affil[2]{Department of Computer Science, Ramakrishna Mission Vivekananda Educational and Research Institute}
\date{}
\maketitle
{\bf Abstract.} One of the commonly used approaches to capture dependence in multivariate survival data is through the frailty variables. The identifiability issues should be carefully investigated while modeling multivariate survival with or without competing risks. The use of non-parametric frailty distribution(s) is sometimes preferred for its robustness and flexibility properties. In this paper, we consider modeling of bivariate survival data with competing risks through four different kinds of non-parametric frailty and parametric baseline cause-specific hazard functions to 
investigate the corresponding model identifiability. We make the common assumption of the frailty mean being equal to unity. 


\section{Introduction}

In a previous work (Ghosh et al, 2024b), we have considered the issue of model identifiability for bivariate failure time data with competing risks assuming non-parametric baseline cause-specific hazard functions. The dependence between the two failure times corresponding to two related individuals (or, a pair of observations from the same individual) are modeled through four different Gamma frailty distributions. These are (1) shared frailty, (2) correlated frailty, (3) shared cause-specific frailty and (4) correalted cause-specific frailty, respectively. In addition to modeling dependence, these frailty components also describe heterogeneity between individuals to some extent. Besides proving identifiability of the corresponding models under fairly general conditions, that work also proved non-identifiability of the model when both the baseline cause-specific hazard functions and the frailty distributions are arbitrary and non-parametric. This leaves us with the other possibility when the baseline cause-specific hazard functions may belong to some parametric family, but the frailty variable has an arbitrary distribution. Therefore, it is relevant to study identifiability of models with parametric hazards but with nonparametric frailties. \\

In this paper, we consider the parametric class of baseline cause-specific hazard functions introduced by Ghosh et al (2024a) which includes hazards forms of Exponential, Weibull, Gamma, and Log-logistic distributions as special cases. This parametric class is particularly helpful as it contains various shapes of hazard functions namely increasing, decreasing, and bathtub-shaped. Hence, it enables us to choose appropriate cause-specific hazard functions according to the nature of observed data. We also consider the four different frailty structures mentioned above. The common assumption of these models to ensure identifiability is the frailty means being equal to unity. \\ 

Note that a typical observation for bivariate survival data with competing risks for a pair of individuals is of the form $(T_1,T_2,J_1,J_2)$, where $T_k$ is the failure time of the $k$th individual of a pair with $J_k$ as the cause of failure, for $k=1,2$. Allowing for possibly different sets of competing risks for the two individuals, we assume that $J_k$ has the support $\{1,\cdots L_k\}$, only one of which is responsible for failure of the $k$th individual, for $k=1,2$.  
Bivariate survival data with competing risks can be modeled through the joint sub-distribution function $F_{j_1j_2}(t_1,t_2)$ defined as
$$F_{j_1j_2}(t_1,t_2) = Pr\big[T_1 \leq t_1, T_2 \leq t_2, J_1 = j_1, J_2 = j_2\big], $$
where $t_{k} > 0, j_{k}=1,\cdots,L_k, k = 1,2$. 
We now introduce frailty random variable(s) possibly depending on the cause of failure. Let us write the frailty vector $\boldsymbol{\epsilon^{(k)}} = (\epsilon^{(k)}_{1},\cdots,\epsilon^{(k)}_{L_k})$, 
where $\epsilon^{(k)}_{j}>0$ denotes the frailty variable  corresponding to  the $j$th cause of failure for the $k$th individual, for $j = 1,\cdots,L_k$ and $k=1,2$. Since dependence between the two failure times $(T_1,T_2)$ with the corresponding causes $(J_1,J_2)$ is modeled through the random frailty terms $\boldsymbol{\epsilon^{(1)}}$ and $\boldsymbol{\epsilon^{(2)}}$, there is independence between $(T_1,J_1)$ and $(T_2,J_2)$ conditional on ($\boldsymbol{\epsilon^{(1)}}, \boldsymbol{\epsilon^{(2)}})$. 
The cause-specific hazard function 
$\lambda^{(k)}_{j}(t_k | \boldsymbol{\epsilon^{(k)}})$ for the failure of the $k$th individual at time $T_{k} = t_k$ due to cause $j$, conditional on the frailty vector $\boldsymbol{\epsilon^{(k)}}$, is defined as 
\begin{equation}\label{EqGeneralmodel}
    \lambda^{(k)}_{j}(t_k | \boldsymbol{\epsilon^{(k)}}) = h^{(k)}_{0j}(t_k)\epsilon^{(k)}_{j}, 
\end{equation}
for $t_{k} > 0, j= 1,\cdots,L_{k}$ and 
$k=1,2$, where $h^{(k)}_{0j}(t_k)$ is the $j$th baseline cause-specific hazard at time $t_k$ for the $k$th individual. \\

Let us define a class of parametric baseline cause-specific hazard functions $h^{(k)}_{0j}(t_k;\boldsymbol{\xi})$ (See Ghosh et al, 2024a), suppressing the dependence on $j$ and $k$, as  
\begin{equation}\label{parametric_family_haz}
    h(t;\gamma,\alpha) = a(\gamma,\alpha)t^{\gamma - 1}b(t;\gamma,\alpha),
\end{equation} 
for all $t > 0$, where $\alpha>0$ and $\gamma>0$ are scale and shape parameters, respectively, $a(\gamma,\alpha)$ is a positive-valued function of $\gamma$ and $\alpha$; it is also assumed that   $a(\gamma,\alpha)$ is an one-to-one function in $\alpha$ for fixed $\gamma$. The function $b(t;\gamma,\alpha)$ is  positive for all $t>0$ such that $\lim\limits_{t \to 0+}b(t;\gamma,\alpha) = 1$. It can be easily proved, letting ${t \to 0+}$, that this parametric class (\ref{parametric_family_haz}) is identifiable in $\alpha$ and $\gamma$ in the sense that $ h(t;\gamma,\alpha)= h(t;\tilde{\gamma},\tilde{\alpha})$ for all $t>0$ implies $\gamma=\tilde{\gamma}$ and $\alpha=\tilde{\alpha}$. 
Note that the parameters $\alpha$ and $\gamma$ depend on $k$ and $j$ in general. Therefore, the parameter $\boldsymbol{\xi}$ is the parameter vector of all these $\alpha$'s and $\gamma$'s. Note that the parameter vector $\boldsymbol{\xi}$ can be written as 
$$\boldsymbol{\xi}=\{(\gamma^{(k)}_j,\alpha^{(k)}_j),\ j=1,\cdots,L_k,\ k=1,2\}.$$
We can also write $\boldsymbol{\xi}= (\boldsymbol{\xi^{(1)}},\boldsymbol{\xi^{(2)}})$, where $\boldsymbol{\xi^{(k)}}=\{(\gamma^{(k)}_j,\alpha^{(k)}_j),\ j=1,\cdots,L_k\}$, for $k=1,2$. With this break-up of notation, clearly, the baseline cause-specific hazard function $h^{(k)}_{0j}(t_k;\boldsymbol{\xi})$ depends on $\boldsymbol{\xi}$ only through $\boldsymbol{\xi^{(k)}_j}=(\gamma^{(k)}_j,\alpha^{(k)}_j)$. 
Let us write $\boldsymbol{\Xi}$ as the space of all these $\boldsymbol{\xi}$'s satisfying the above conditions leading to the parametric class of baseline cause-specific hazard functions under study. As noted in Ghosh et al (2024a), this class includes expressions of \textit{Exponential, Weibull, Gamma, Log-Logistic} hazard functions as special cases. \\

As in Ghosh et al (2024a,b), the conditional survival function of the $k$th individual, given the frailty vector $\boldsymbol{\epsilon^{(k)}}$, is
$$
 S^{(k)}(t_k;\boldsymbol{\xi^{(k)}}|\boldsymbol{\epsilon^{(k)}}) 
    = \exp{\bigg[-\sum\limits_{j=1}^{L_k} \epsilon^{(k)}_{j} 
    H^{(k)}_{0j}(t_k;\boldsymbol{\xi^{(k)}})\bigg]},
$$
where $H^{(k)}_{0j}(t_k;\boldsymbol{\xi^{(k)}}) = \int\limits_{0}^{t_{k}}h^{(k)}_{0j}(u_k;\boldsymbol{\xi^{(k)}})du_k$,  and the $j$th sub-distribution function of  the $k$th individual, conditional on the frailty vector $\boldsymbol{\epsilon^{(k)}}$, is 
$$
   F_j^{(k)}(t_{k};\boldsymbol{\xi^{(k)}} | \boldsymbol{\epsilon^{(k)}})  
   = \int\limits_{0}^{t_k}h^{(k)}_{0j}(u_k;\boldsymbol{\xi^{(k)}})  \epsilon^{(k)}_{j}\exp{\Bigg[-\sum\limits_{j'= 1}^{L_k}H^{(k)}_{0j'}(u_k;\boldsymbol{\xi^{(k)}})  \epsilon^{(k)}_{j'}\Bigg]}du_k, 
$$
for all $j = 1,\cdots,L_{k}$ and $k = 1,2$. 
Therefore, using conditional independence given frailty, the joint unconditional sub-distribution function $F_{j_1j_2}(t_1,t_2;\boldsymbol{\xi})$ under the general model (\ref{EqGeneralmodel}) is
\begin{align}\label{eq:JointModel}
    F_{j_1j_2}(t_1,t_2;\boldsymbol{\xi})
    &= \int\limits_{0}^{\infty}\cdots\int\limits_{0}^{\infty}\Bigg[\int\limits_{0}^{t_1}\int\limits_{0}^{t_2}
    \prod_{k=1}^2 \left(h^{(k)}_{0j_{k}}(u_k;\boldsymbol{\xi^{(k)}}) \epsilon^{(k)}_{j_k}\right) \exp{\Bigg(-\sum_{k=1}^2 
    \sum\limits_{j= 1}^{L_k}H^{(k)}_{0j}(u_k;\boldsymbol{\xi^{(k)}})  \epsilon^{(k)}_{j}\Bigg)}du_2 du_1\Bigg] \nonumber\\ 
    &\qquad\qquad\qquad\qquad\qquad\times dG(\boldsymbol{\epsilon^{(1)}},\boldsymbol{\epsilon^{(2)}}),
\end{align}
for all $j_{k} = 1,\cdots,L_{k}$ and $k = 1,2$, where  $G(\boldsymbol{\epsilon^{(1)}},\boldsymbol{\epsilon^{(2)}})$ is an arbitrary probability measure on $[0,\infty)^{L_1}\times [0,\infty)^{L_2}$, representing the joint frailty distribution of  $(\boldsymbol{\epsilon^{(1)}},\boldsymbol{\epsilon^{(2)}})$. 
In this work, we 
assume it to belong to a family $\mathcal{G}$ of arbitrary joint distribution functions. It is to be noted that this joint sub-distribution function also depends on the joint frailty distribution $G( \boldsymbol{\epsilon^{(1)}}, \boldsymbol{\epsilon^{(2)}})\in \mathcal{G}$; so we write this as $F_{j_1j_2}(t_1,t_2;\boldsymbol{\xi},G)$ for completeness. 
The focus of this paper is to investigate identifiability of the model for bivariate failure time with competing risks $(T_1,T_2,J_1,J_2)$ given by the joint sub-distribution function 
$F_{j_{1}j_{2}}\big(t_1,t_2;\boldsymbol{\xi},G\big)$ with the four different arbitrary frailty distributions, as mentioned earlier. For this, we need the following definition of model identifiability. 
\begin{definition}
   The model (\ref{EqGeneralmodel}) for bivariate failure time with competing risks with the baseline cause-specific hazards given by (\ref{parametric_family_haz}) 
   is identifiable 
   within $\boldsymbol{\Xi}\times\mathcal{G}$ 
   if, for some $\boldsymbol{\xi},\ \boldsymbol{\tilde{\xi}}\in\ \boldsymbol{\Xi}$ and $G,\ \tilde{G}\in\ \mathcal{G}$, the equality 
$$F_{j_{1}j_{2}}\big(t_1,t_2;\boldsymbol{\xi},G\big) = F_{j_{1}j_{2}}\big(t_1,t_2;\boldsymbol{\tilde{\xi}},\tilde{G}\big), $$
for all $t_{k} > 0,\ j_{k} = 1,\cdots,L_{k}$ and $k=1,2$, implies $\boldsymbol{\xi}=\boldsymbol{\tilde{\xi}}$ and $G( \boldsymbol{\epsilon^{(1)}}, \boldsymbol{\epsilon^{(2)}})=\tilde{G}( \boldsymbol{\epsilon^{(1)}}, \boldsymbol{\epsilon^{(2)}})$ almost surely. 
\end{definition}

In the Sections 2-5, we consider the four different types of frailty distributions, namely, (1) shared frailty, (2) correlated frailty, (3) shared cause-specific frailty and (4) correalted cause-specific frailty, respectively, to study identifiability of the corresponding models. Note that the family $\mathcal{G}$ of the joint frailty distribution functions differs in nature and dimension for the four different frailty types. Section 6 ends with some concluding remarks. 

\section{Non-parametric shared frailty model} 

The non-parametric shared frailty model is the simplest one with one common frailty variable $\epsilon$ that is shared between the two individuals in a pair. So, we have $\epsilon^{(k)}_{j}=\epsilon$, for all $j=1,\cdots,L_k$ and $k=1,2$. This model enjoys the flexibility of allowing two different sets of competing risks for the two individuals. Formally, the model is given by 
\begin{equation}\label{Parametric_hazard_Nonpara_shared_frailty}
    \lambda^{(k)}_{j}(t_k;\boldsymbol{\xi^{(k)}_{j}} | \epsilon) = h^{(k)}_{0j}(t_k;\boldsymbol{\xi^{(k)}_{j}})\epsilon, 
\end{equation}
for $t_k > 0,\ j = 1,\cdots,L_{k}$ and $k = 1,2$. Here the shared frailty $\epsilon>0$ is assumed to have an arbitrary distribution function $G$ defined on $[0,\infty)$. So, the class $\mathcal{G}$ of the arbitrary frailty distributions is given by 
$$\mathcal{G}=\{G(\epsilon): \epsilon>0\}.$$ 

As in Section 1, the conditional survival function of $k$th individual, given the shared frailty $\epsilon$, is 
$$
    S^{(k)}(t_k ; \boldsymbol{\xi^{(k)}} | \epsilon) = \exp{\bigg[-\epsilon\sum\limits_{j = 1}^{L_k} H^{(k)}_{0j}(t_k;\boldsymbol{\xi^{(k)}}) \bigg]}
    = \exp{\bigg[- \epsilon H^{(k)}_{0}(t_k;\boldsymbol{\xi^{(k)}})\bigg]},$$
where $H^{(k)}_{0}(t_k;\boldsymbol{\xi^{(k)}})= \sum\limits_{j = 1}^{L_k} H^{(k)}_{0j}(t_k;\boldsymbol{\xi^{(k)}})$, 
for all $t_{k} > 0$ and $k = 1,2$. Also, the conditional $j$th sub-distribution function of $k$th individual, given $\epsilon$,  is
\begin{align*}
\begin{aligned}
F^{(k)}_{j}(t_k;\boldsymbol{\xi^{(k)}}| \epsilon) 
&= \int\limits_{0}^{t_k}\epsilon h^{(k)}_{0j}(u_k;\boldsymbol{\xi^{(k)}_{j}}) \exp{\Big[-\epsilon H^{(k)}_{0}(u_k;\boldsymbol{\xi^{(k)}})\Big]}du_k\\
&= \int\limits_{0}^{t_k}\epsilon 
a(\gamma^{(k)}_{j},\alpha^{(k)}_{j})u_k^{\gamma^{(k)}_{j} - 1}b(u_k;\gamma^{(k)}_{j},\alpha^{(k)}_{j}) \exp{\Big[-\epsilon H^{(k)}_{0}(u_k;\boldsymbol{\xi^{(k)}})\Big]}du_k,
\end{aligned}
\end{align*}
using (\ref{parametric_family_haz}), for all $t_{k} > 0,\ j = 1,\cdots,L_{k}$ and $k = 1,2$. As in (\ref{eq:JointModel}), 
the unconditional joint sub-distribution function $F_{j_1j_2}(t_1,t_2;\boldsymbol{\xi},G)$ is given by $$
\int\limits_{0}^{\infty}\int\limits_{0}^{t_1}\int\limits_{0}^{t_2}\epsilon^{2}\prod\limits_{k=1}^{2}\Bigg(a(\gamma^{(k)}_{j_k},\alpha^{(k)}_{j_k})u_k^{\gamma^{(k)}_{j_k} - 1}b(u_{k};\gamma^{(k)}_{j_k},\alpha^{(k)}_{j_k})\Bigg) 
    \times\exp{\bigg[-\epsilon \sum\limits_{k=1}^{2}H^{(k)}_{0}(u_k;\boldsymbol{\xi^{(k)}})\bigg]}du_2du_1 dG(\epsilon).$$
Similarly, the unconditional $j$th sub-distribution function for $k$th individual is given by 
$$
   F^{(k)}_{j}(t_k;\boldsymbol{\xi^{(k)}},G) =  \int\limits_{0}^{\infty}\int\limits_{0}^{t_k}\epsilon a(\gamma^{(k)}_{j},\alpha^{(k)}_{j}) u_k^{\gamma^{(k)}_{j} - 1}b(u_k;\gamma^{(k)}_{j},\alpha^{(k)}_{j}) \times 
   \exp{\bigg[-\epsilon H^{(k)}_{0}(u_k;\boldsymbol{\xi^{(k)}})\bigg]}du_{k}dG(\epsilon),$$
with the corresponding unconditional sub-density function 
\begin{equation}
    \label{eq:subdensity_SharedFrialty}
   f^{(k)}_{j}(t_k;\boldsymbol{\xi^{(k)}},G) =  \int\limits_{0}^{\infty}\epsilon a(\gamma^{(k)}_{j},\alpha^{(k)}_{j}) t_k^{\gamma^{(k)}_{j} - 1}b(t_k;\gamma^{(k)}_{j},\alpha^{(k)}_{j}) \times 
   \exp{\bigg[-\epsilon H^{(k)}_{0}(t_k;\boldsymbol{\xi^{(k)}})\bigg]}dG(\epsilon).
   \end{equation}

\begin{definition}
The non-parametric shared frailty model (\ref{Parametric_hazard_Nonpara_shared_frailty}) for bivariate failure time with competing risks with the baseline cause-specific hazards given by (\ref{parametric_family_haz}) 
is identifiable within $\boldsymbol{\Xi}\times\mathcal{G}$ 
   if, for some $\boldsymbol{\xi},\ \boldsymbol{\tilde{\xi}}\in\ \boldsymbol{\Xi}$ and $G,\ \tilde{G}\in\ \mathcal{G}=\{G(\epsilon)\}$, as defined in the beginning of this section,  the equality 
$$F_{j_{1}j_{2}}\big(t_1,t_2;\boldsymbol{\xi},G\big) = F_{j_{1}j_{2}}\big(t_1,t_2;\boldsymbol{\tilde{\xi}},\tilde{G}\big), $$
for all $t_{k} > 0,\ j_{k} = 1,\cdots,L_{k}$ and $k=1,2$, implies $\boldsymbol{\xi}=\boldsymbol{\tilde{\xi}}$ and $G(\epsilon)=\tilde{G}(\epsilon)$ almost surely. 
\end{definition}
\begin{theorem}
 The nonparametric shared frailty model
 (\ref{Parametric_hazard_Nonpara_shared_frailty}) for bivariate failure time with competing risks with the baseline cause-specific hazards given by (\ref{parametric_family_haz}) 
is identifiable within $\boldsymbol{\Xi}\times\mathcal{G}$, 
 provided $\lim_{t\rightarrow\infty}H^{(k)}_{0}(t;\boldsymbol{\xi^{(k)}}) = \infty$ for all $\boldsymbol{\xi^{(k)}}$ and  
 $\mathbb{E}(\epsilon) = \int_0^{\infty}\epsilon dG(\epsilon)=1$. 
\end{theorem}

\begin{proof}
From the equality of the joint unconditional sub-distribution functions as in {\bf Definition 2}, for all $t_k >
0,\ j_k = 1,\cdots,L_k$ and $k = 1, 2$, we get equality of the unconditional sub-distribution functions $F^{(k)}_{j}(t_k;\boldsymbol{\xi^{(k)}},G)$ and $F^{(k)}_{j}(t_k;\boldsymbol{\tilde{\xi}^{(k)}},\tilde{G})$; in particular, we get equality of the corresponding unconditional sub-density functions $f^{(k)}_{j}(t_k;\boldsymbol{\xi^{(k)}},G)$ and $f^{(k)}_{j}(t_k;\boldsymbol{\tilde{\xi}^{(k)}},\tilde{G})$. From the expression above in (\ref{eq:subdensity_SharedFrialty}), we have 
\begin{gather*}
\int\limits_{0}^{\infty}\epsilon a(\gamma^{(k)}_{j},\alpha^{(k)}_{j})
t_k^{\gamma^{(k)}_{j}-1} b(t_k;\gamma^{(k)}_{j},\alpha^{(k)}_{j}) \exp{[-\epsilon  H^{(k)}_{0}(t_k;\boldsymbol{\xi^{(k)}})]}dG(\epsilon)\nonumber \\ 
= \int\limits_{0}^{\infty}\epsilon a(\tilde{\gamma}^{(k)}_{j},\tilde{\alpha}^{(k)}_{j})
t_k^{\tilde{\gamma}^{(k)}_{j}-1} b(t_k;\tilde{\gamma}^{(k)}_{j},\tilde{\alpha}^{(k)}_{j}) \exp{[-\epsilon  H^{(k)}_{0}(t_k;\boldsymbol{\tilde{\xi}^{(k)}})]}d\tilde{G}(\epsilon)
\end{gather*}
for all $t_{k} > 0,\ j = 1,\cdots,L_{k}$ and $k=1,2$. \\

Following the technique used by Heckman and Singer (1984) to show identifiability of  shape parameters, we rearrange the above equation to write 
\begin{equation}
\label{Parametric_hazard_Nonparametric_shared_frailty_Eq2}
t^{\gamma^{(k)}_{j} - \tilde{\gamma}^{(k)}_{j}}_{k} 
\frac{a(\gamma^{(k)}_{j},\alpha^{(k)}_{j})b(t_k;\gamma^{(k)}_{j},\alpha^{(k)}_{j})A(t_k;\boldsymbol{\xi^{(k)}},G)} {a(\tilde{\gamma}^{(k)}_{j},\tilde{\alpha}^{(k)}_{j})b(t_k;\tilde{\gamma}^{(k)}_{j},\tilde{\alpha}^{(k)}_{j})A(t_k;\boldsymbol{\tilde{\xi}^{(k)}},\tilde{G})}
= 1, 
\end{equation}
where 
$A(t_k;\boldsymbol{\xi^{(k)}},G) = \int\limits_{0}^{\infty}\epsilon \exp{[-\epsilon H^{(k)}_{0}(t_k;\boldsymbol{\xi^{(k)}})]}dG(\epsilon)$, for $t_k>0,\ j=1,\cdots,L_k$ and $k=1,2$. 
Note that, by dominated convergence theorem, we have $\lim\limits_{t_k \to 0+}A(t_k;\boldsymbol{\xi^{(k)}},G) = \int\limits_{0}^{\infty}\epsilon dG(\epsilon) = 1$ by the assumption of the theorem. Similarly, we have $\lim\limits_{t_k \to 0+}A(t_k;\boldsymbol{\tilde{\xi}^{(k)}},\tilde{G}) = \int\limits_{0}^{\infty} \epsilon d\tilde{G}(\epsilon) = 1$. \\

Now, depending on whether $\gamma^{(k)}_{j} > \tilde{\gamma}^{(k)}_{j}$ or $\gamma^{(k)}_{j} < \tilde{\gamma}^{(k)}_{j}$, the limit of the left hand side of (\ref{Parametric_hazard_Nonparametric_shared_frailty_Eq2}), as $t_k \to 0+$, is 0 or $\infty$, while the same limit of the right hand side of (\ref{Parametric_hazard_Nonparametric_shared_frailty_Eq2}) is 1, which is a contradiction. Therefore, we have 
$\gamma^{(k)}_{j} = \tilde{\gamma}^{(k)}_{j}$, 
for $j=1,\cdots,L_k$ and $k=1,2$. Using this equality in (\ref{Parametric_hazard_Nonparametric_shared_frailty_Eq2}) and re-arranging the terms again, we get 
$$\frac{a(\gamma^{(k)}_{j},\tilde{\alpha}^{(k)}_{j})}{a(\gamma^{(k)}_{j},\alpha^{(k)}_{j})}=
\frac{b(t_k;\gamma^{(k)}_{j},\alpha^{(k)}_{j})A(t_k;\boldsymbol{\xi^{(k)}},G)} {b(t_k;\tilde{\gamma}^{(k)}_{j},\tilde{\alpha}^{(k)}_{j})A(t_k;\boldsymbol{\tilde{\xi}^{(k)}},\tilde{G})}.$$
Now, as before, letting $t_k \to 0+$ in both sides of the above equation, we have $$\frac{a(\gamma^{(k)}_{j},\tilde{\alpha}^{(k)}_{j})}{a(\gamma^{(k)}_{j},\alpha^{(k)}_{j})}=1.$$
Therefore, using the one-to-one property of the function $a(\cdot,\cdot)$, we get $\alpha^{(k)}_{j}=\tilde{\alpha}^{(k)}_{j}$, for $j=,\cdots,L_k$ and $k=1,2$. Therefore, we now have $\boldsymbol{\xi}= \boldsymbol{\tilde{\xi}}$. \\

Now, equality of the joint unconditional sub-distribution functions also implies equality of the unconditional survival functions 
$S^{(k)}(t_k;\boldsymbol{\xi^{(k)}},G) = S^{(k)}(t_k;\boldsymbol{\tilde{\xi}^{(k)}},\tilde{G})$, with $\boldsymbol{\xi^{(k)}}= \boldsymbol{\tilde{\xi}^{(k)}}$, for all $t_{k} > 0$. Using the expression of the conditional survival function $S^{(k)}(t_k;\boldsymbol{\xi^{(k)}}|\epsilon)$ in the beginning of this section, we therefore have 
$$\int\limits_{0}^{\infty}\exp{\Big[-\epsilon H^{(k)}_{0}(t_k;\boldsymbol{\xi^{(k)}})\Big]}dG(\epsilon) 
= \int\limits_{0}^{\infty}\exp{\Big[- \epsilon H^{(k)}_{0}(t_k;\boldsymbol{\xi^{(k)}})\Big]}d\tilde{G}(w),$$
for all  $t_{k} > 0$. 
Since the cumulative baseline hazard functions are monotonically increasing with $H^{(k)}_{0}(0;\boldsymbol{\xi^{(k)}}) = 0$ and 
$\lim_{t\rightarrow\infty}H^{(k)}_{0}(t;\boldsymbol{\xi^{(k)}}) = \infty$, there exists exactly one time point $t^{\ast}_{kn}$ such that $H^{(k)}_{0}(t^{\ast}_{kn};\boldsymbol{\xi^{(k)}}) = n$ for each $n \in \mathbb{N}$, by intermediate value property. Now, taking limit as $t_{k} \to t^{\ast}_{kn}$ for each $n \in \mathbb{N}$ in the above equation, we get 
\begin{equation*}
\int\limits_{0}^{\infty}\exp{(-n\epsilon)}dG(\epsilon) 
= \int\limits_{0}^{\infty}\exp{(-n\epsilon)}d\tilde{G}(\epsilon) \, \, \text{for each}\, \, n \in \mathbb{N}. 
\end{equation*}
From the above equation, it follows that 
$G(\epsilon) = \tilde{G}(\epsilon)$ almost surely for all $ \epsilon > 0$ 
by the uniqueness theorem regarding Laplace-Stieltjes  transformation (See Lin and Dou, 2021). Hence, the identifiability of this model is proved.

\end{proof}

\section{Non-parametric correlated frailty model}

Let $\epsilon^{(1)}>0$ and $\epsilon^{(2)}>0$ be two correlated frailty variables associated with failyre times of the first and the second individual, respectively, regardless of the causes of failure. Dependence between the two failure times is modeled through the correlation between these two frailty variables. As in the previous section, the two individuals may be allowed to have different sets of competing risks. Let us denote their joint distribution function as $G(\epsilon^{(1)},\epsilon^{(2)})$ defined on $[0,\infty)\times [0,\infty)$, so that the family $\mathcal{G}$ is now defined as 
$$\mathcal{G}=\{G(\epsilon^{(1)},\epsilon^{(2)}): \epsilon^{(1)}>0,\epsilon^{(2)}>0\}.$$
Let us denote the marginal distribution of $\epsilon^{(k)}$ as $G^{(k)}(\epsilon^{(k)})$, for $k=1,2$, which can be obtained as $G^{(1)}(\epsilon^{(1)})=G(\epsilon^{(1)},\infty)$ and $G^{(2)}(\epsilon^{(2)})=G(\infty,\epsilon^{(2)})$. 
Formally, the non-parametric correlated frailty model is 
\begin{equation}
\label{Parametric_hazard_Nonpara_correlated_frailty}
 \lambda^{(k)}_{j}(t_k;\boldsymbol{\xi^{(k)}_j} | \epsilon^{(k)}) = h^{(k)}_{0j}(t_k;\boldsymbol{\xi^{(k)}_j})\epsilon^{(k)},
\end{equation}
for $t_{k} > 0,\ j = 1,\cdots,L_k$ and $k = 1,2$. 
As before, the conditional survival function for the $k$th individual, given  $\epsilon^{(k)}$, is 
$$
S^{(k)}(t_k;\boldsymbol{\xi^{(k)}} | \epsilon^{(k)}) 
=  \exp{\Bigg[- \epsilon^{(k)}H^{(k)}_{0}\big(t_k;\boldsymbol{\xi^{(k)}}\big)\Bigg]}
$$
and the conditional $j$th sub-distribution function of the $k$th individual, given   $\epsilon^{(k)}$,  is
$$
F^{(k)}_{j}(t_k;\boldsymbol{\xi^{(k)}} | \epsilon^{(k)}) =  \int\limits_{0}^{t_k}\epsilon^{(k)} a(\gamma^{(k)}_{j}, \alpha^{(k)}_{j}) u_k^{\gamma^{(k)}_{j} - 1}b(u_k;\gamma^{(k)}_{j},\alpha^{(k)}_{j}) \exp{[-\epsilon^{(k)}H^{(k)}_{0}(u_k;\boldsymbol{\xi^{(k)}})]}du_k,$$
for all $t_{k} > 0,\ j = 1,\cdots,L_{k}$ and $k = 1,2$. 
Therefore, the unconditional joint sub-distribution function $F_{j_1j_2}(t_1,t_2;\boldsymbol{\xi},G)$ is given by 
$$
\int\limits_{0}^{\infty}\int\limits_{0}^{\infty}\int\limits_{0}^{t_1}\int\limits_{0}^{t_2}\prod\limits_{k=1}^{2}\Bigg(\epsilon^{(k)} a(\gamma^{(k)}_{j_k},\alpha^{(k)}_{j_k}) u^{\gamma^{(k)}_{j_k} - 1}_{k} b(u_{k};\gamma^{(k)}_{j_k},\alpha^{(k)}_{j_k})\Bigg)
\exp{\Bigg[-\sum\limits_{k=1}^{2} \epsilon^{(k)}H^{(k)}_{0}(u_k;\boldsymbol{\xi^{(k)}})\Bigg]}du_2du_1 dG(\epsilon^{(1)},\epsilon^{(2)}),$$
for all $t_{k} > 0,\ j_{k} = 1,\cdots,L_{k}$ and $k = 1,2$. 
Similarly, the unconditional $j$th sub-distribution function $F^{(k)}_{j}(t_k;\boldsymbol{\xi^{(k)}},G)$ of the $k$th individual is 
$$
\int\limits_{0}^{\infty}\int\limits_{0}^{t_k}\epsilon^{(k)} a(\gamma^{(k)}_{j}, \alpha^{(k)}_{j}) u_k^{\gamma^{(k)}_{j} - 1}b(u_k;\gamma^{(k)}_{j},\alpha^{(k)}_{j}) \exp{[-\epsilon^{(k)}H^{(k)}_{0}(u_k;\boldsymbol{\xi^{(k)}})]}du_kdG^{(k)}(\epsilon^{(k)})$$
with the corresponding unconditional sub-density  function $f^{(k)}_{j}(t_k;\boldsymbol{\xi^{(k)}},G)$ given by 
$$
\int\limits_{0}^{\infty}\epsilon^{(k)} a(\gamma^{(k)}_{j}, \alpha^{(k)}_{j}) t_k^{\gamma^{(k)}_{j} - 1}b(t_k;\gamma^{(k)}_{j},\alpha^{(k)}_{j}) \exp{[-\epsilon^{(k)}H^{(k)}_{0}(t_k;\boldsymbol{\xi^{(k)}})]}dG^{(k)}(\epsilon^{(k)}),$$
for all $t_{k} > 0,\ j = 1,\cdots,L_{k}$ and $k = 1,2$. \\

\begin{definition}
The non-parametric correlated frailty model (\ref{Parametric_hazard_Nonpara_correlated_frailty}) 
for bivariate failure time with competing risks with the baseline cause-specific hazards given by (\ref{parametric_family_haz}) 
is identifiable within $\boldsymbol{\Xi}\times\mathcal{G}$ 
   if, for some $\boldsymbol{\xi},\ \boldsymbol{\tilde{\xi}}\in\ \boldsymbol{\Xi}$ and $G,\ \tilde{G}\in\ \mathcal{G}=\{G(\epsilon^{(1)},\epsilon^{(2)})\}$, as defined in the beginning of this section, the equality 
$$F_{j_{1}j_{2}}\big(t_1,t_2;\boldsymbol{\xi},G\big) = F_{j_{1}j_{2}}\big(t_1,t_2;\boldsymbol{\tilde{\xi}},\tilde{G}\big), $$
for all $t_{k} > 0,\ j_{k} = 1,\cdots,L_{k}$ and $k=1,2$, implies $\boldsymbol{\xi}=\boldsymbol{\tilde{\xi}}$ and $G(\epsilon^{(1)},\epsilon^{(2)})=\tilde{G}(\epsilon^{(1)},\epsilon^{(2)})$ almost surely. 
\end{definition}
\begin{theorem}
 The non-parametric correlated frailty model
 (\ref{Parametric_hazard_Nonpara_correlated_frailty}) for bivariate failure time with competing risks with the baseline cause-specific hazards given by (\ref{parametric_family_haz}) 
is identifiable within $\boldsymbol{\Xi}\times\mathcal{G}$, 
 provided $\lim_{t\rightarrow\infty}H^{(k)}_{0}(t;\boldsymbol{\xi^{(k)}}) = \infty$ for all $\boldsymbol{\xi^{(k)}}$ and  
 $\mathbb{E}(\epsilon^{(k)}) = \int_0^{\infty}\epsilon^{(k)} dG^{(k)}(\epsilon^{(k)})=1$, for $k=1,2$. 
\end{theorem}
\begin{proof}
As in the previous section, equating the unconditional $j$th 
sub-density functions $f^{(k)}_{j}(t_k;\boldsymbol{\xi^{(k)}},G)$ and $f^{(k)}_{j}(t_k;\boldsymbol{\tilde{\xi}^{(k)}},\tilde{G})$, we get  
\begin{gather*}
\int\limits_{0}^{\infty}\epsilon^{(k)} a(\gamma^{(k)}_{j},\alpha^{(k)}_{j})
t_k^{\gamma^{(k)}_{j}-1} b(t_k;\gamma^{(k)}_{j},\alpha^{(k)}_{j}) \exp{[-\epsilon^{(k)}  H^{(k)}_{0}(t_k;\boldsymbol{\xi^{(k)}})]}dG^{(k)}(\epsilon^{(k)})\nonumber \\ 
\qquad\qquad = \int\limits_{0}^{\infty}\epsilon^{(k)} a(\tilde{\gamma}^{(k)}_{j},\tilde{\alpha}^{(k)}_{j})
t_k^{\tilde{\gamma}^{(k)}_{j}-1} b(t_k;\tilde{\gamma}^{(k)}_{j},\tilde{\alpha}^{(k)}_{j}) \exp{[-\epsilon^{(k)}  H^{(k)}_{0}(t_k;\boldsymbol{\tilde{\xi}^{(k)}})]}d\tilde{G}^{(k)}(\epsilon^{(k)}),
\end{gather*}
for all $t_{k} > 0,\ j = 1,\cdots,L_{k}$ and $k=1,2$. Re-arranging the terms as in the previous section, we now have 
$$
t^{\gamma^{(k)}_{j} - \tilde{\gamma}^{(k)}_{j}}_{k} 
\frac{a(\gamma^{(k)}_{j},\alpha^{(k)}_{j})b(t_k;\gamma^{(k)}_{j},\alpha^{(k)}_{j})A(t_k;\boldsymbol{\xi^{(k)}},G^{(k)})} {a(\tilde{\gamma}^{(k)}_{j},\tilde{\alpha}^{(k)}_{j})b(t_k;\tilde{\gamma}^{(k)}_{j},\tilde{\alpha}^{(k)}_{j})A(t_k;\boldsymbol{\tilde{\xi}^{(k)}},\tilde{G}^{(k)})}
= 1,$$ 
for $t_{k} > 0,\ j = 1,\cdots,L_{k}$ and $k=1,2$. Note that this is the same identity as that in (\ref{Parametric_hazard_Nonparametric_shared_frailty_Eq2}) in the previous section with $G$ replaced by $G^{(k)}$. Therefore, following the same arguments as those used in {\bf Theorem 2.1} and using the assumption of Theorem 3.1, we can prove that  $\boldsymbol{\xi}= \boldsymbol{\tilde{\xi}}$. \\

Now, the equality of the unconditional joint sub-distribution functions implies equality of the unconditional joint survival functions $S(t_1,t_2;\boldsymbol{\xi},G)$ and $S(t_1,t_2;\boldsymbol{\tilde{\xi}},\tilde{G})$ with $\boldsymbol{\xi}= \boldsymbol{\tilde{\xi}}$, for all $t_k>0$ and $k=1,2$. Using the expression of the conditional survival functions $S^{(k)}(t_k;\boldsymbol{\xi^{(k)}} | \epsilon^{(k)})$, for $k=1,2$, in the beginning of this section and using  conditional independence, the above equality of the joint survival functions gives 
$$
\int\limits_{0}^{\infty}\int\limits_{0}^{\infty}\exp{\Big[-\sum\limits_{k=1}^{2} \epsilon^{(k)} H^{(k)}_{0}(t_k;\boldsymbol{\xi^{(k)}})\Big]}dG(\epsilon^{(1)}, \epsilon^{(2)}) 
= \int\limits_{0}^{\infty}\int\limits_{0}^{\infty}\exp{\Big[-\sum\limits_{k=1}^{2} \epsilon^{(k)} H^{(k)}_{0}(t_k;\boldsymbol{\xi^{(k)}})\Big]}d\tilde{G}(\epsilon^{(1)}, \epsilon^{(2)}).$$
Since the cumulative baseline hazard functions are monotonically increasing with $H^{(k)}_{0}(0;\boldsymbol{\xi^{(k)}})=0$ and $\lim_{t\rightarrow\infty}H^{(k)}_{0}(t;\boldsymbol{\xi^{(k)}}) = \infty$, 
there exists exactly one point $t^{\ast}_{kn}$ such that $H^{(k)}_{0}(t^{\ast}_{kn};\boldsymbol{\xi^{(k)}})=n$, for every $n \in \mathbb{N}$ and for $k = 1,2$, by intermediate value property. Therefore, taking limit as $t_{k} \to t^{\ast}_{kn}$ for every $n \in \mathbb{N}$ and for $k = 1,2$ in the above equation, we get 
$$
\int\limits_{0}^{\infty}\int\limits_{0}^{\infty}\exp{[-\sum\limits_{k=1}^{2} \epsilon^{(k)}n]}dG(\epsilon^{(1)}, \epsilon^{(2)})
= 
\int\limits_{0}^{\infty}\int\limits_{0}^{\infty}\exp{[-\sum\limits_{k=1}^{2} \epsilon^{(k)}n]}d\tilde{G}(\epsilon^{(1)}, \epsilon^{(2)}).$$
Therefore, it follows from the uniqueness theorem regarding bivariate Laplace-Stieltjes transformation (See Lin and Dou, 2021) that
$G(\epsilon^{(1)}, \epsilon^{(2)}) = \tilde{G}(\epsilon^{(1)}, \epsilon^{(2)})$ 
almost surely for $\epsilon^{(k)}>0$, for $k=1,2$. Hence, the identifiability of this model is proved.
\end{proof}

\section{Non-parametric shared cause-specific frailty model}

Unlike the previous two sections, we now assume the non-parametric frailty to depend on the cause of failure, but shared by the two individuals in a pair (See Ghosh et al., 2024a,b). In this case, we need the set of competing risks to be the same for both the individuals since the shared frailty is cause-specific. So, we assume $L_1=L_2=L$, say. 
Let $\epsilon_j>0$ denote the shared frailty for the $j$th cause, for $j=1,\cdots,L$, and write $\boldsymbol{\epsilon}=(\epsilon_1,\cdots,\epsilon_L)$.  
Dependence between the two individuals is modeled through the common frailty variables in $\boldsymbol{\epsilon}$. 
The model is given by 
\begin{equation}\label{Parametric_hazard_Nonapara_shared_cause_specific_frailty}
 \lambda^{(k)}_{j}(t_k;\boldsymbol{\xi^{(k)}_{j}} | \boldsymbol{\epsilon}) = h^{(k)}_{0j}(t_k;\boldsymbol{\xi^{(k)}_{j}})\epsilon_{j},
\end{equation}
for $t_k > 0,\ j = 1,\cdots,L$ and $k = 1,2$. Let us write the joint distribution function of $\boldsymbol{\epsilon}$ as $G(\boldsymbol{\epsilon})$ and the corresponding marginal distribution of $\epsilon_j$ as $G_j(\epsilon_j)$, for $j=1,\cdots,L$, which can be obtained from $G(\boldsymbol{\epsilon})$. In contrast with the Gamma shared cause-specific frailty model of Ghosh et al. (2024a,b), we do not need to assume independence between the $\epsilon_j$'s. So, the family of $\mathcal{G}$ of frailty distributions is defined as 
$$\mathcal{G}=\{G(\boldsymbol{\epsilon}): \epsilon_j>0,\ j=1,\cdots,L\}.$$

As before, the conditional survival function for the $k$th individual, given $\boldsymbol{\epsilon}$, is 
$$
S^{(k)}(t_k;\boldsymbol{\xi^{(k)}} |\boldsymbol{\epsilon}) 
=  \exp{\Bigg[- \sum_{j=1}^L \epsilon_j H^{(k)}_{0j}\big(t_k;\boldsymbol{\xi^{(k)}_j}\big)\Bigg]}
$$
and the conditional $j$th sub-distribution function of the $k$th individual, given $\boldsymbol{\epsilon}$,  is
$$
F^{(k)}_{j}(t_k;\boldsymbol{\xi^{(k)}} | \boldsymbol{\epsilon}) =  \int\limits_{0}^{t_k}\epsilon_j a(\gamma^{(k)}_{j}, \alpha^{(k)}_{j}) u_k^{\gamma^{(k)}_{j} - 1}b(u_k;\gamma^{(k)}_{j},\alpha^{(k)}_{j}) \exp{[- \sum_{j'=1}^L \epsilon_{j'} H^{(k)}_{0j'}\big(t_k;\boldsymbol{\xi^{(k)}_{j'}}\big)
]}du_k,$$
for all $t_{k} > 0,\ j = 1,\cdots,L$ and $k = 1,2$. 
Therefore, the unconditional $j$th sub-distribution function $F^{(k)}_{j}(t_k;\boldsymbol{\xi^{(k)}},G)$ of the $k$th individual is 
$$
\int\limits_{0}^{\infty}\cdots\int\limits_{0}^{\infty}
\int\limits_{0}^{t_k}\epsilon_j a(\gamma^{(k)}_{j}, \alpha^{(k)}_{j}) u_k^{\gamma^{(k)}_{j} - 1}b(u_k;\gamma^{(k)}_{j},\alpha^{(k)}_{j}) \exp{[- \sum_{j'=1}^L \epsilon_{j'} H^{(k)}_{0j'} \big(u_k;\boldsymbol{\xi^{(k)}_{j'}}\big)]}du_k dG(\boldsymbol{\epsilon})$$
with the corresponding unconditional sub-density  function $f^{(k)}_{j}(t_k;\boldsymbol{\xi^{(k)}},G)$ given by 
$$\int\limits_{0}^{\infty}\cdots\int\limits_{0}^{\infty}
\epsilon_j a(\gamma^{(k)}_{j}, \alpha^{(k)}_{j}) t_k^{\gamma^{(k)}_{j} - 1}b(t_k;\gamma^{(k)}_{j},\alpha^{(k)}_{j}) \exp{[- \sum_{j'=1}^L \epsilon_{j'} H^{(k)}_{0j'} \big(t_k;\boldsymbol{\xi^{(k)}_{j'}}\big)]} dG(\boldsymbol{\epsilon})$$
for all $t_{k} > 0,\ j = 1,\cdots,L$ and $k = 1,2$. Similarly, the unconditional joint sub-distribution function $F_{j_1j_2}(t_1,t_2;\boldsymbol{\xi},G)$ is 
$$
\int\limits_{0}^{\infty}\cdots\int\limits_{0}^{\infty} \int\limits_{0}^{t_1}\int\limits_{0}^{t_2} \prod\limits_{k=1}^{2} \Bigg(\epsilon_{j_k} a(\gamma^{(k)}_{j_k},\alpha^{(k)}_{j_k}) u^{\gamma^{(k)}_{j_k} - 1}_{k} b(u_{k};\gamma^{(k)}_{j_k},\alpha^{(k)}_{j_k})\Bigg)
    \exp{\Bigg[-\sum\limits_{k=1}^{2} \sum_{j=1}^L \epsilon_{j}H^{(k)}_{0j}(u_k;\boldsymbol{\xi^{(k)}_j})\Bigg]}du_2du_1 dG(\boldsymbol{\epsilon}),$$
and the corresponding unconditional joint sub-density  function $f_{j_1,j_2}(t_1,t_2;\boldsymbol{\xi},G)$ is 
$$
\int\limits_{0}^{\infty}\cdots\int\limits_{0}^{\infty} \prod\limits_{k=1}^{2} \Bigg(\epsilon_{j_k} a(\gamma^{(k)}_{j_k},\alpha^{(k)}_{j_k}) t^{\gamma^{(k)}_{j_k} - 1}_{k} b(t_{k};\gamma^{(k)}_{j_k},\alpha^{(k)}_{j_k})\Bigg)
    \exp{\Bigg[-\sum\limits_{k=1}^{2} \sum_{j=1}^L \epsilon_{j}H^{(k)}_{0j}(t_k;\boldsymbol{\xi^{(k)}_j}) \Bigg]} dG(\boldsymbol{\epsilon}),$$
for $t_{k} > 0,\ j_{k} = 1,\cdots,L$ and $k = 1,2$. \\

\begin{definition}
The non-parametric shared cause-specific frailty model (\ref{Parametric_hazard_Nonapara_shared_cause_specific_frailty}) for bivariate failure time with competing risks with the baseline cause-specific hazards given by (\ref{parametric_family_haz}) 
is identifiable within $\boldsymbol{\Xi}\times\mathcal{G}$ 
   if, for some $\boldsymbol{\xi},\ \boldsymbol{\tilde{\xi}}\in\ \boldsymbol{\Xi}$ and $G,\ \tilde{G}\in\ \mathcal{G}=\{G(\boldsymbol{\epsilon})\}$, as defined in the beginning of this section, the equality 
$$F_{j_{1}j_{2}}\big(t_1,t_2;\boldsymbol{\xi},G\big) = F_{j_{1}j_{2}}\big(t_1,t_2;\boldsymbol{\tilde{\xi}},\tilde{G}\big), $$
for all $t_{k} > 0,\ j_{k} = 1,\cdots,L$ and $k=1,2$, implies $\boldsymbol{\xi}=\boldsymbol{\tilde{\xi}}$ and $G(\boldsymbol{\epsilon})=\tilde{G}(\boldsymbol{\epsilon})$ almost surely. 
\end{definition}
\begin{theorem}
 The non-parametric shared cause-specific frailty model (\ref{Parametric_hazard_Nonapara_shared_cause_specific_frailty}) 
  for bivariate failure time with competing risks with the baseline cause-specific hazards given by (\ref{parametric_family_haz}) 
is identifiable within $\boldsymbol{\Xi}\times\mathcal{G}$, 
 provided $\lim_{t\rightarrow\infty}H^{(k)}_{0j}(t;\boldsymbol{\xi^{(k)}_j}) = \infty$, for all $\boldsymbol{\xi^{(k)}_j}$, and  
 $\mathbb{E}(\epsilon_j) = \int_0^{\infty}\epsilon_j dG_j(\epsilon_j)=1$, for $j=1,\cdots,L$ and $k=1,2$. 
\end{theorem}
\begin{proof}
As before, equating the unconditional $j$th 
sub-density functions $f^{(k)}_{j}(t_k;\boldsymbol{\xi^{(k)}},G)$ and $f^{(k)}_{j}(t_k;\boldsymbol{\tilde{\xi}^{(k)}},\tilde{G})$, we get 
\begin{align*}
&\int\limits_{0}^{\infty}\cdots\int\limits_{0}^{\infty}
\epsilon_j a(\gamma^{(k)}_{j}, \alpha^{(k)}_{j}) t_k^{\gamma^{(k)}_{j} - 1}b(t_k;\gamma^{(k)}_{j},\alpha^{(k)}_{j}) \exp{[- \sum_{j'=1}^L \epsilon_{j'} H^{(k)}_{0j'} \big(t_k;\boldsymbol{\xi^{(k)}_{j'}}\big)]} dG(\boldsymbol{\epsilon}) = \\
&\qquad \int\limits_{0}^{\infty}\cdots\int\limits_{0}^{\infty}
\epsilon_j a(\tilde{\gamma}^{(k)}_{j}, \tilde{\alpha}^{(k)}_{j}) t_k^{\tilde{\gamma}^{(k)}_{j} - 1} b(t_k;\tilde{\gamma}^{(k)}_{j},\tilde{\alpha}^{(k)}_{j}) \exp{[- \sum_{j'=1}^L \epsilon_{j'} H^{(k)}_{0j'} \big(t_k;\boldsymbol{\tilde{\xi}^{(k)}_{j'}}\big)]} d\tilde{G}(\boldsymbol{\epsilon}),
\end{align*}
for all $t_{k} > 0,\ j = 1,\cdots,L$ and $k=1,2$. Re-arranging the terms as in the previous section, we now have 
$$
t^{\gamma^{(k)}_{j} - \tilde{\gamma}^{(k)}_{j}}_{k} 
\frac{a(\gamma^{(k)}_{j},\alpha^{(k)}_{j})b(t_k;\gamma^{(k)}_{j},\alpha^{(k)}_{j})B_j(t_k;\boldsymbol{\xi^{(k)}},G)} {a(\tilde{\gamma}^{(k)}_{j},\tilde{\alpha}^{(k)}_{j})b(t_k;\tilde{\gamma}^{(k)}_{j},\tilde{\alpha}^{(k)}_{j})B_j(t_k;\boldsymbol{\tilde{\xi}^{(k)}},\tilde{G})} = 1,$$ 
where $B_j(t_k;\boldsymbol{\xi^{(k)}},G)= \int\limits_{0}^{\infty}\cdots\int\limits_{0}^{\infty}
\epsilon_j \exp{[- \sum_{j'=1}^L \epsilon_{j'} H^{(k)}_{0j'} \big(t_k;\boldsymbol{\xi^{(k)}_{j'}}\big)]} dG(\boldsymbol{\epsilon})$, for all $t_{k} > 0,\ j = 1,\cdots,L$ and $k=1,2$. Note that this is the same identity as that in (\ref{Parametric_hazard_Nonparametric_shared_frailty_Eq2}) in Section 2 with the univariate $G$ replaced by the $L$-variate $G$ and $A(t_k;\boldsymbol{\xi^{(k)}},G)$ replaced by $B_j(t_k;\boldsymbol{\xi^{(k)}},G)$. However, as for $A(t_k;\boldsymbol{\xi^{(k)}},G)$, one can prove, by using the 
dominated convergence theorem and the assumption of Theorem 4.1, that $$\lim\limits_{t_k \to 0+}B_j(t_k;\boldsymbol{\xi^{(k)}},G)= \lim\limits_{t_k \to 0+}B_j(t_k;\boldsymbol{\tilde{\xi}^{(k)}},\tilde{G})=1.$$
Therefore, following the same arguments as those used in {\bf Theorem 2.1} and using the assumption of Theorem 4.1, we can prove that  $\boldsymbol{\xi}= \boldsymbol{\tilde{\xi}}$. \\ 

Now, the equality of the unconditional joint sub-distribution functions implies equality of the unconditional joint survival functions $S(t_1,t_2;\boldsymbol{\xi},G)$ and $S(t_1,t_2;\boldsymbol{\tilde{\xi}},\tilde{G})$ with $\boldsymbol{\xi}= \boldsymbol{\tilde{\xi}}$, for all $t_k>0$ and $k=1,2$. Using the expression of the conditional survival functions $S^{(k)}(t_k;\boldsymbol{\xi^{(k)}} | \boldsymbol{\epsilon})$, for $k=1,2$, in the beginning of this section and using conditional independence, the above equality of the unconditional joint survival functions gives 
$$
\int\limits_{0}^{\infty}\cdots\int\limits_{0}^{\infty} \exp{\Big[- \sum_{j=1}^L\epsilon_j \sum\limits_{k=1}^{2} H^{(k)}_{0j}(t_k;\boldsymbol{\xi^{(k)}_j})\Big]}dG(\boldsymbol{\epsilon}) 
= \int\limits_{0}^{\infty}\cdots\int\limits_{0}^{\infty}\exp{\Big[-
\sum_{j=1}^L\epsilon_j \sum\limits_{k=1}^{2} H^{(k)}_{0j}(t_k;\boldsymbol{\xi^{(k)}_j})\Big]}d\tilde{G}(\boldsymbol{\epsilon})
.$$

Now, let $\{m_{1n}\}$ be a sequence of positive, strictly increasing, real numbers and bounded above strictly by 1, for $n \in \mathbb{N}$. 
Since the cumulative baseline cause-specific hazard functions are monotonically increasing with $H^{(k)}_{0j}(0;\boldsymbol{\xi^{(k)}_j})=0$ and $\lim_{t\rightarrow\infty}H^{(k)}_{0j}(t;\boldsymbol{\xi^{(k)}_j}) = \infty$, for all $\boldsymbol{\xi^{(k)}_j}$ and for $j=1,\cdots,L$, 
there exists exactly one point $t^{\ast}_{kn}$ such that $H^{(k)}_{01}(t^{\ast}_{kn};\boldsymbol{\xi^{(k)}_1})=m_{1n}$, for every $n \in \mathbb{N}$ and for $k = 1,2$, by intermediate value property. So, we can write $t^{\ast}_{kn}=H^{-(k)}_{01}(m_{1n};\boldsymbol{\xi^{(k)}_1})$, where $H^{-(k)}_{01}(\cdot;\boldsymbol{\xi^{(k)}_1})$ denotes the inverse function of $H^{(k)}_{01}(\cdot;\boldsymbol{\xi^{(k)}_1})$, for $k=1,2$, which exist by the definition in  (\ref{parametric_family_haz}). Let us define, for $j=2,\cdots,L$ and for every $n \in \mathbb{N}$, 
$$m_{jn} = \sum_{k=1}^2 H^{(k)}_{0j}(t^{\ast}_{kn};\boldsymbol{\xi^{(k)}_j}) = 
\sum_{k=1}^2 H^{(k)}_{0j}(H^{-(k)}_{01}(m_{1n};\boldsymbol{\xi^{(k)}_1});\boldsymbol{\xi^{(k)}_j}).$$ These $m_{jn}$'s, for fixed $j$,  are clearly a sequence of positive and strictly increasing real numbers. Also, since $m_{1n}<1$ for every $n \in \mathbb{N}$, we have 
$$m_{jn} 
< 
\sum_{k=1}^2 H^{(k)}_{0j}(H^{-(k)}_{01}(1;\boldsymbol{\xi^{(k)}_1});\boldsymbol{\xi^{(k)}_j})=c_j,\quad\mbox{ say,}$$ 
for $j=1,\cdots,L$ and for every $n \in \mathbb{N}$. Therefore, we have 
$$\sum\limits_{n=1}^{\infty}\frac{1}{m_{jn}}= \infty,$$
for $j=1,\cdots,L$. Then, taking limit as $t_{k} \to t^{\ast}_{kn}$ for every $n \in \mathbb{N}$ and for $k = 1,2$ in the above equality of the joint survival functions, we get 
$$
\int\limits_{0}^{\infty}\cdots\int\limits_{0}^{\infty}\exp{\Big[-\sum\limits_{j=1}^{L}\epsilon_jm_{jn}\Big]} dG(\boldsymbol{\epsilon}) = \int\limits_{0}^{\infty}\cdots\int\limits_{0}^{\infty}\exp{\Big[-\sum\limits_{j=1}^{L}\epsilon_{j}m_{jn}\Big]}d\tilde{G}(\boldsymbol{\epsilon}),$$
for every $n \in \mathbb{N}$. Hence $G(\boldsymbol{\epsilon}) = \tilde{G}(\boldsymbol{\epsilon})$ almost surely for $\boldsymbol{\epsilon} > \boldsymbol{0}$, 
by the uniqueness theorem regarding multivariate Laplace-Stieltjes transformation (See Lin and Dou, 2021). Therefore, the identifiability of this model is proved. 

\end{proof}

\section{Non-parametric correlated cause-specific frailty model}

This is a further generalization of the frailty models we considered in the previous three sections. As in Section 3, we have two correlated frailty variables for the two individuals; however, as in Section 4, these two frailties depend on the specific cause of failure. That is, for each cause, say the $j$th, there is a pair of frailty variables $\boldsymbol{\epsilon_j}=(\epsilon_j^{(1)}, \epsilon_j^{(2)})$ for the two individuals which are correlated as in Section 3 (See Ghosh et al., 2024a,b). 
In this case also, we need the set of competing risks to be the same for both the individuals since the pair of frailties is cause-specific. So, we have $L_1=L_2=L$. 
Let us write $\boldsymbol{\epsilon^{(k)}}=(\epsilon_1^{(k)},\cdots,\epsilon_L^{(k)})$ denoting the frailties for  the $k$th individual, for $k=1,2$. 
Dependence between the two individuals is modeled through the pairs $\boldsymbol{\epsilon_j}$'s for $j=1,\cdots,L$. Formally, the model is given by 
\begin{equation}\label{Parametric_hazard_Nonapara_correlated_cause_specific_frailty}
 \lambda^{(k)}_{j}(t_k;\boldsymbol{\xi^{(k)}_{j}} | \boldsymbol{\epsilon^{(k)}}) = h^{(k)}_{0j}(t_k;\boldsymbol{\xi^{(k)}_{j}})\epsilon_{j}^{(k)},
\end{equation}
for $t_k > 0,\ j = 1,\cdots,L$ and $k = 1,2$. Let us write the joint distribution function of $\boldsymbol{\epsilon}=(\boldsymbol{\epsilon^{(1)}},\boldsymbol{\epsilon^{(2)}})$ as $G(\boldsymbol{\epsilon})$ 
and the corresponding marginal distribution of the frailty vector $\boldsymbol{\epsilon^{(k)}}$ as $G^{(k)}(\boldsymbol{\epsilon^{(k)}})$, for $k=1,2$, which can be obtained from $G(\boldsymbol{\epsilon})$. For that matter, let $G^{(k)}_j(\epsilon^{(k)}_j)$ denote the marginal distribution of the frailty variable $\epsilon^{(k)}_j$, for all $j$ and $k$. 
Here also, we do not need to assume independence between the $\epsilon_j^{(k)}$'s for different $j$ and $k$. So, the family of $\mathcal{G}$ of frailty distributions is defined as 
$$\mathcal{G}=\{G(\boldsymbol{\epsilon}): \epsilon_j^{(k)}>0,\ j=1,\cdots,L,\ k=1,2\}.$$

The conditional survival function for the $k$th individual, given the frailty vector $\boldsymbol{\epsilon^{(k)}}$, is 
$$
S^{(k)}(t_k;\boldsymbol{\xi^{(k)}} |\boldsymbol{\epsilon^{(k)}}) 
=  \exp{\Bigg[- \sum_{j=1}^L \epsilon_j^{(k)} H^{(k)}_{0j}\big(t_k;\boldsymbol{\xi^{(k)}_j}\big)\Bigg]}
$$
and the conditional $j$th sub-distribution function of the $k$th individual, given $\boldsymbol{\epsilon^{(k)}}$,  is
$$
F^{(k)}_{j}(t_k;\boldsymbol{\xi^{(k)}} | \boldsymbol{\epsilon^{(k)}}) =  \int\limits_{0}^{t_k}\epsilon_j^{(k)} a(\gamma^{(k)}_{j}, \alpha^{(k)}_{j}) u_k^{\gamma^{(k)}_{j} - 1}b(u_k;\gamma^{(k)}_{j},\alpha^{(k)}_{j}) \exp{[- \sum_{j'=1}^L \epsilon_{j'}^{(k)} H^{(k)}_{0j'}\big(t_k;\boldsymbol{\xi^{(k)}_{j'}}\big)
]}du_k,$$
for all $t_{k} > 0,\ j = 1,\cdots,L$ and $k = 1,2$. 
Therefore, the unconditional $j$th sub-distribution function $F^{(k)}_{j}(t_k;\boldsymbol{\xi^{(k)}},G)$ of the $k$th individual is 
$$
\int\limits_{0}^{\infty}\cdots\int\limits_{0}^{\infty}
\int\limits_{0}^{t_k}\epsilon_j^{(k)} a(\gamma^{(k)}_{j}, \alpha^{(k)}_{j}) u_k^{\gamma^{(k)}_{j} - 1}b(u_k;\gamma^{(k)}_{j},\alpha^{(k)}_{j}) \exp{[- \sum_{j'=1}^L \epsilon_{j'}^{(k)} H^{(k)}_{0j'} \big(u_k;\boldsymbol{\xi^{(k)}_{j'}}\big)]}du_k dG^{(k)}(\boldsymbol{\epsilon^{(k)}})$$
with the corresponding unconditional sub-density  function $f^{(k)}_{j}(t_k;\boldsymbol{\xi^{(k)}},G)$ given by 
$$\int\limits_{0}^{\infty}\cdots\int\limits_{0}^{\infty}
\epsilon_j^{(k)} a(\gamma^{(k)}_{j}, \alpha^{(k)}_{j}) t_k^{\gamma^{(k)}_{j} - 1}b(t_k;\gamma^{(k)}_{j},\alpha^{(k)}_{j}) \exp{[- \sum_{j'=1}^L \epsilon_{j'}^{(k)} H^{(k)}_{0j'} \big(t_k;\boldsymbol{\xi^{(k)}_{j'}}\big)]} dG^{(k)}(\boldsymbol{\epsilon^{(k)}}),$$
for all $t_{k} > 0,\ j = 1,\cdots,L$ and $k = 1,2$. 
Similarly, the unconditional joint sub-distribution function $F_{j_1j_2}(t_1,t_2;\boldsymbol{\xi},G)$ is 
$$
\int\limits_{0}^{\infty}\cdots\int\limits_{0}^{\infty} \int\limits_{0}^{t_1}\int\limits_{0}^{t_2} \prod\limits_{k=1}^{2} \Bigg(\epsilon_{j_k}^{(k)} a(\gamma^{(k)}_{j_k},\alpha^{(k)}_{j_k}) u^{\gamma^{(k)}_{j_k} - 1}_{k} b(u_{k};\gamma^{(k)}_{j_k},\alpha^{(k)}_{j_k})\Bigg)
    \exp{\Bigg[-\sum\limits_{k=1}^{2} \sum_{j=1}^L \epsilon_{j}^{(k)}H^{(k)}_{0j}(u_k;\boldsymbol{\xi^{(k)}_j})\Bigg]}du_2du_1 dG(\boldsymbol{\epsilon}),$$
and the corresponding unconditional joint sub-density  function $f_{j_1j_2}(t_1,t_2;\boldsymbol{\xi},G)$ is 
$$
\int\limits_{0}^{\infty}\cdots\int\limits_{0}^{\infty} \prod\limits_{k=1}^{2} \Bigg(\epsilon_{j_k}^{(k)} a(\gamma^{(k)}_{j_k},\alpha^{(k)}_{j_k}) t^{\gamma^{(k)}_{j_k} - 1}_{k} b(t_{k};\gamma^{(k)}_{j_k},\alpha^{(k)}_{j_k})\Bigg)
    \exp{\Bigg[-\sum\limits_{k=1}^{2} \sum_{j=1}^L \epsilon_{j}^{(k)}H^{(k)}_{0j}(t_k;\boldsymbol{\xi^{(k)}_j}) \Bigg]} dG(\boldsymbol{\epsilon}),$$
for $t_{k} > 0,\ j_{k} = 1,\cdots,L$ and $k = 1,2$. \\

\begin{definition}
The non-parametric correlated cause-specific frailty model (\ref{Parametric_hazard_Nonapara_correlated_cause_specific_frailty}) for bivariate failure time with competing risks with the baseline cause-specific hazards given by (\ref{parametric_family_haz}) 
is identifiable within $\boldsymbol{\Xi}\times\mathcal{G}$ 
   if, for some $\boldsymbol{\xi},\ \boldsymbol{\tilde{\xi}}\in\ \boldsymbol{\Xi}$ and $G,\ \tilde{G}\in\ \mathcal{G}=\{G(\boldsymbol{\epsilon})\}$, as defined in the beginning of this section, the equality 
$$F_{j_{1}j_{2}}\big(t_1,t_2;\boldsymbol{\xi},G\big) = F_{j_{1}j_{2}}\big(t_1,t_2;\boldsymbol{\tilde{\xi}},\tilde{G}\big), $$
for all $t_{k} > 0,\ j_{k} = 1,\cdots,L$ and $k=1,2$, implies $\boldsymbol{\xi}=\boldsymbol{\tilde{\xi}}$ and $G(\boldsymbol{\epsilon})=\tilde{G}(\boldsymbol{\epsilon})$ almost surely. 
\end{definition}
\begin{theorem}
The non-parametric correlated cause-specific frailty model (\ref{Parametric_hazard_Nonapara_correlated_cause_specific_frailty}) 
for bivariate failure time with competing risks with the baseline cause-specific hazards given by (\ref{parametric_family_haz}) 
is identifiable within $\boldsymbol{\Xi}\times\mathcal{G}$, 
 provided $\lim_{t\rightarrow\infty}H^{(k)}_{0j}(t;\boldsymbol{\xi^{(k)}_j}) = \infty$, for all $\boldsymbol{\xi^{(k)}_j}$, and  
 $\mathbb{E}(\epsilon_j^{(k)}) = \int_0^{\infty}\epsilon_j^{(k)} dG_j^{(k)}(\epsilon_j^{(k)})=1$, for $j=1,\cdots,L$ and $k=1,2$. 
\end{theorem}
\begin{proof}
As in the previous sections, equating the unconditional $j$th 
sub-density functions $f^{(k)}_{j}(t_k;\boldsymbol{\xi^{(k)}},G)$ and $f^{(k)}_{j}(t_k;\boldsymbol{\tilde{\xi}^{(k)}},\tilde{G})$, we get  
\begin{gather*}
\int\limits_{0}^{\infty}\cdots\int\limits_{0}^{\infty}
\epsilon_j^{(k)} a(\gamma^{(k)}_{j}, \alpha^{(k)}_{j}) t_k^{\gamma^{(k)}_{j} - 1}b(t_k;\gamma^{(k)}_{j},\alpha^{(k)}_{j}) \exp{[- \sum_{j'=1}^L \epsilon_{j'}^{(k)} H^{(k)}_{0j'} \big(t_k;\boldsymbol{\xi^{(k)}_{j'}}\big)]} dG^{(k)}(\boldsymbol{\epsilon^{(k)}}) \nonumber \\ 
\qquad\qquad =
\int\limits_{0}^{\infty}\cdots\int\limits_{0}^{\infty}
\epsilon_j^{(k)} a(\tilde{\gamma}^{(k)}_{j}, \tilde{\alpha}^{(k)}_{j}) t_k^{\tilde{\gamma}^{(k)}_{j} - 1}b(t_k;\tilde{\gamma}^{(k)}_{j},\tilde{\alpha}^{(k)}_{j}) \exp{[- \sum_{j'=1}^L \epsilon_{j'}^{(k)} H^{(k)}_{0j'} \big(t_k;\boldsymbol{\tilde{\xi}^{(k)}_{j'}}\big)]} d\tilde{G}^{(k)}(\boldsymbol{\epsilon^{(k)}}), 
\end{gather*}
for all $t_{k} > 0,\ j = 1,\cdots,L$ and $k=1,2$. 
Re-arranging the terms as in the previous section, we can write  
$$
t^{\gamma^{(k)}_{j} - \tilde{\gamma}^{(k)}_{j}}_{k} 
\frac{a(\gamma^{(k)}_{j},\alpha^{(k)}_{j})b(t_k;\gamma^{(k)}_{j},\alpha^{(k)}_{j})C_{jk}(t_k;\boldsymbol{\xi^{(k)}},G^{(k)})} {a(\tilde{\gamma}^{(k)}_{j},\tilde{\alpha}^{(k)}_{j})b(t_k;\tilde{\gamma}^{(k)}_{j},\tilde{\alpha}^{(k)}_{j})C_{jk}(t_k;\boldsymbol{\tilde{\xi}^{(k)}},\tilde{G}^{(k)})} = 1,$$ 
where $C_{jk}(t_k;\boldsymbol{\xi^{(k)}},G^{(k)})= \int\limits_{0}^{\infty}\cdots\int\limits_{0}^{\infty}
\epsilon_j^{(k)} \exp{[- \sum_{j'=1}^L \epsilon_{j'}^{(k)} H^{(k)}_{0j'} \big(t_k;\boldsymbol{\xi^{(k)}_{j'}}\big)]} dG^{(k)}(\boldsymbol{\epsilon^{(k)}})$, for all $t_{k} > 0,\ j = 1,\cdots,L$ and $k=1,2$. This is again the same identity as that in (\ref{Parametric_hazard_Nonparametric_shared_frailty_Eq2}) in Section 2 with the univariate $G$ replaced by the $L$-variate $G^{(k)}$ and $A(t_k;\boldsymbol{\xi^{(k)}},G)$ replaced by $C_{jk}(t_k;\boldsymbol{\xi^{(k)}},G^{(k)})$. However, as for $A(t_k;\boldsymbol{\xi^{(k)}},G)$, one can prove, by using the 
dominated convergence theorem and the assumption of Theorem 5.1, that $$\lim\limits_{t_k \to 0+}C_{jk}(t_k;\boldsymbol{\xi^{(k)}},G^{(k)})= \lim\limits_{t_k \to 0+}C_{jk}(t_k;\boldsymbol{\tilde{\xi}^{(k)}},\tilde{G}^{(k)})=1.$$
Therefore, following the same arguments as those used in {\bf Theorem 2.1} and using the assumption of Theorem 5.1, we can prove that  $\boldsymbol{\xi}= \boldsymbol{\tilde{\xi}}$. \\ 

Equality of the unconditional joint sub-distribution functions implies equality of the unconditional joint survival functions $S(t_1,t_2;\boldsymbol{\xi},G)$ and $S(t_1,t_2;\boldsymbol{\tilde{\xi}},\tilde{G})$ with $\boldsymbol{\xi}= \boldsymbol{\tilde{\xi}}$, for all $t_k>0$ and $k=1,2$. Using the expression of the conditional survival functions $S^{(k)}(t_k;\boldsymbol{\xi^{(k)}} | \boldsymbol{\epsilon^{(k)}})$, for $k=1,2$, in the beginning of this section and using conditional independence, the above equality of the joint survival functions gives 
$$
\int\limits_{0}^{\infty}\cdots\int\limits_{0}^{\infty} \exp{\Big[- \sum_{j=1}^L \sum\limits_{k=1}^{2} \epsilon_j^{(k)} H^{(k)}_{0j}(t_k;\boldsymbol{\xi^{(k)}_j})\Big]}dG(\boldsymbol{\epsilon}) 
= \int\limits_{0}^{\infty}\cdots\int\limits_{0}^{\infty} \exp{\Big[-
\sum_{j=1}^L\sum\limits_{k=1}^{2}\epsilon_j^{(k)} H^{(k)}_{0j}(t_k;\boldsymbol{\xi^{(k)}_j})\Big]}d\tilde{G}(\boldsymbol{\epsilon}).$$

As in the previous section, let $\{m_{1n}^{(k)}\}$ be a sequence of positive, strictly increasing, real numbers and bounded above strictly by 1, for $k=1,2$ and $n \in \mathbb{N}$.  
Since the cumulative baseline cause-specific hazard functions are monotonically increasing with $H^{(k)}_{0j}(0;\boldsymbol{\xi^{(k)}_j})=0$ and $\lim_{t\rightarrow\infty}H^{(k)}_{0j}(t;\boldsymbol{\xi^{(k)}_j}) = \infty$, for $j=1,\cdots,L$, 
there exists exactly one point $t^{\ast}_{kn}$ such that $H^{(k)}_{01}(t^{\ast}_{kn};\boldsymbol{\xi^{(k)}_1})=m_{1n}^{(k)}$, for every $n \in \mathbb{N}$ and for $k = 1,2$, by intermediate value property. For $k=1,2$, let us define  
$$m_{jn}^{(k)} =  H^{(k)}_{0j}(t^{\ast}_{kn};\boldsymbol{\xi^{(k)}_j}) = H^{(k)}_{0j}(H^{-(k)}_{01}(m_{1n}^{(k)};\boldsymbol{\xi^{(k)}_1});\boldsymbol{\xi^{(k)}_j}),$$ 
for $j=2,\cdots,L$ and $n \in \mathbb{N}$. 
These $m_{jn}^{(k)}$'s, for fixed $j$ and $k$,  are clearly a sequence of positive and strictly increasing real numbers. Also, since $m_{1n}^{(k)}<1$ for every $n \in \mathbb{N}$, we have 
$$m_{jn}^{(k)} < H^{(k)}_{0j}(H^{-(k)}_{01}(1;\boldsymbol{\xi^{(k)}_1});\boldsymbol{\xi^{(k)}_j})=d_j^{(k)},\quad\mbox{ say,}$$ 
for $j=1,\cdots,L$ and for every $n \in \mathbb{N}$. Therefore, we have 
$$\sum\limits_{n=1}^{\infty}\frac{1}{m_{jn}^{(k)}}= \infty,$$
for $j=1,\cdots,L$ and $k=1,2$. Then, taking limit as $t_{k} \to t^{\ast}_{kn}$ for every $n \in \mathbb{N}$ and for $k = 1,2$ in the above equality of the joint survival functions, we get 
$$
\int\limits_{0}^{\infty}\cdots\int\limits_{0}^{\infty} \exp{\Big[- \sum_{j=1}^L \sum\limits_{k=1}^{2} \epsilon_j^{(k)} m_{jn}^{(k)}
\Big]}dG(\boldsymbol{\epsilon}) 
= \int\limits_{0}^{\infty}\cdots\int\limits_{0}^{\infty} \exp{\Big[-
\sum_{j=1}^L\sum\limits_{k=1}^{2}\epsilon_j^{(k)} m_{jn}^{(k)} \Big]}d\tilde{G}(\boldsymbol{\epsilon}),$$
for every $n \in \mathbb{N}$. Hence $G(\boldsymbol{\epsilon}) = \tilde{G}(\boldsymbol{\epsilon})$ almost surely for $\boldsymbol{\epsilon} > \boldsymbol{0}$, 
by the uniqueness theorem regarding multivariate Laplace-Stieltjes transformation (See Lin and Dou, 2021). Therefore, the identifiability of this model is proved. 

\end{proof}

\section{Concluding Remarks}

Modeling of bivariate failure time data with competing risks using frailty has been studied by several authors, for example, see Bandeen-Roche and Liang ($2002$) and Gorfine and Hsu ($2011$). Even though identifiability study of the resulting models is an essential aspect of statistical inference, there is no investigation for such identifiability, as far as we know. 
In this work, we have studied this model identifiability with a particular class of parametric baseline cause-specific hazard functions and four different types of non-parametric frailty distributions. We have proved identifiability results under fairly reasonable assumptions. Identifiability of models with this particular class of baseline cause-specific hazard functions with different kinds of Gamma frailty models has been shown by Ghosh et al. (2024a), which also follows from the results of the present paper. \\ 

In this paper, we have only studied the models with a parametric class of baseline cause-specific hazard functions which includes a number of commonly used popular hazard functions as special cases. However, there are several other choices (for example, Gumbel and generalized F) for baseline cause-specific of hazard functions. The identifiability study of these models is also of interest which we plan to take up in future. Similarly, extension of the present work to the models for multivariate failure time with competing risks with or without the presence of covariates is of importance. Although the four types of frailty models considered in this work cover a wide range of dependence structure including heterogeneity between individuals, one can possibly think of other forms of frailty, for example, time-dependent frailty. This will certainly mean a big challenge in terms of notation and proof of identifiability as well. 

\section{Refrences}
Bandeen‐Roche, K., and Liang, K. Y. (2002). Modelling multivariate failure time associations in the presence of a competing risk. Biometrika, 89(2), 299-314.\\
Gorfine, M., and Hsu, L. (2011). Frailty-based competing risks model for multivariate survival data. Biometrics, 67(2), 415-426.\\
Ghosh, B., Dewanji, A., and Das, S. (2024a). Parametric Analysis of Bivariate Current Status data with Competing risks using Frailty model. https://doi.org/10.48550/arXiv.2405.05773.\\
Ghosh, B., Dewanji, A., and Das, S. (2024b). Model Identifiability for Bivariate Failure Time Data with Competing Risks: Non-parametric Cause-specific Hazards and Gamma Frailty. Technical report, ASU/2024/02, Applied Statistics Unit, Indian Statistical Institute, 2024.\\ 
Heckman, J., and Singer, B. (1984). The identifiability of the proportional hazard model. The Review of Economic Studies, 51(2), 231-241.\\
Lin, G. D., and Dou, X. (2021). An identity for two integral transforms applied to the uniqueness of a distribution via its laplace-stieljes transform. Statistics, 55(2), 367-385.
\end{document}